\documentclass[12pt]{article}
\usepackage{psfrag}
\usepackage{amsmath}
\usepackage{amsthm}
\usepackage{amssymb}
\usepackage{epsfig}
\usepackage{euscript}
\usepackage{array}
\usepackage{cancel}
\usepackage{mathtools}
\usepackage{empheq}
\usepackage{graphicx}
\usepackage{subfigure}
\usepackage{upgreek}
\usepackage{epstopdf}
\usepackage{booktabs}
\usepackage{mathrsfs}
\usepackage{amsbsy}

\usepackage{hyperref}
\usepackage{verbatim}

\theoremstyle{definition}

\theoremstyle{remark}

\newcounter{multieqs}

\newcommand{\be}{\begin{equation}}
\newcommand{\ee}{\end{equation}}
\newcommand{\eq}[1]{(\ref{#1})}
\newcommand{\bit}{\begin{itemize}}  \newcommand{\eit}{\end{itemize}}
\newcommand{\ben}{\begin{enumerate}}  \newcommand{\een}{\end{enumerate}}

\newcommand{\bra}[1]{\langle #1|}
\newcommand{\ket}[1]{|#1 \rangle}

\newcommand{\xpv}[1]{\langle #1  \rangle}

\newcommand{\bm}[1]{\mbox{\boldmath $#1$}}
\newcommand{\rf}[1]{(\ref{#1})}

\def\bd{\begin{document}}
\def\ed{\end{document}}
\def\bea{\begin{eqnarray}}
\def\eea{\end{eqnarray}}
\let\bm=\bibitem

\def\la{\langle}
\def\ra{\rangle}

\def\npb#1#2#3{Nucl. Phys. {\bf{B#1}} #3 (#2)}
\def\plb#1#2#3{Phys. Lett. {\bf{#1B}} #3 (#2)}
\def\prl#1#2#3{Phys. Rev. Lett. {\bf{#1}} #3 (#2)}
\def\prd#1#2#3{Phys. Rev. {D bf{#1}} #3 (#2)}
\def\cmp#1#2#3{Comm. Math. Phys. {\bf{#1}} #3 (#2)}
\def\cqg#1#2#3{Class. Quantum Grav. {\bf{#1}} #3 (#2)}
\def\nppsa#1#2#3{Nucl. Phys. B (Proc. Suppl.) {\bf{#1A}}#3 (#2)}
\def\ap#1#2#3{Ann. of Phys. {\bf{#1}} #3 (#2)}
\def\ijmp#1#2#3{Int. J. Mod. Phys. {\bf{A#1}} #3 (#2)}
\def\rmp#1#2#3{Rev. Mod. Phys. {\bf{#1}} #3 (#2)}
\def\mpla#1#2#3{Mod. Phys. Lett. {\bf A#1} #3 (#2)}
\def\jhep#1#2#3{J. High Energy Phys. {\bf #1} #3 (#2)}
\def\atmp#1#2#3{Adv. Theor. Math. Phys. {\bf #1} #3 (#2)}

\def\N{{\cal N}}
\def\sst{\scriptscriptstyle}
\def\thetabar{\bar\theta}
\def\Tr{{\rm Tr}}
\def\one{\mbox{1 \kern-.59em {\rm l}}}

\def\a{\alpha}      \def\da{{\dot\alpha}}  \def\dA{{\dot A}}
\def\b{\beta}       \def\db{{\dot\beta}}  
\def\g{\gamma}  \def\G{\Gamma}  \def\dc{{\dot\gamma}}  
\def\d{\delta}  \def\D{\Delta}  \def\ddt{\dot\delta}  
\def\e{\epsilon}        
\def\ve{\varepsilon}  
\def\uve{\upvarepsilon}
\def\f{\phi}    \def\F{\Phi}    \def\vvf{\f}  
\def\h{\eta}  
\def\k{\kappa}  
\def\l{\lambda} \def\L{\Lambda}  
\def\m{\mu} \def\n{\nu}  
\def\o{\omega}  
\def\p{\pi} \def\P{\Pi}  
\def\r{\rho}  
\def\s{\sigma}  \def\S{\Sigma}  
\def\t{\tau}  
\def\th{\theta} \def\Th{\Theta} \def\vth{\vartheta}  
\def\X{\Xeta}  
\def\z{\zeta}  

\def\na{\nabla}  

\def\cA{{\mathscr A}} \def\cB{{\cal B}} \def\cC{{\cal C}}  
\def\cD{{\cal D}} \def\cE{{\cal E}} \def\cF{{\cal F}}  
\def\cG{{\cal G}} \def\cH{{\cal H}} \def\cI{{\cal I}}  
\def\cJ{{\mathscr J}} \def\cK{{\cal K}} \def\cL{{\cal L}}  
\def\cM{{\cal M}} \def\cN{{\cal N}} \def\cO{{\cal O}}  
\def\cP{{\cal P}} \def\cQ{{\cal Q}} \def\cR{{\cal R}}  
\def\cS{{\cal S}} \def\cT{{\cal T}} \def\cU{{\cal U}}  
\def\cV{{\cal V}} \def\cW{{\cal W}} \def\cX{{\cal X}}  
\def\cY{{\cal Y}} \def\cZ{{\cal Z}}

\def\ua{\underline{\alpha}}  
\def\uc{\underline{\phantom{\alpha}}\!\!\!\gamma}  
\def\um{\underline{\mu}}  
\def\ud{\underline\delta}  
\def\ue{\underline\epsilon}  
\def\una{\underline a}\def\unA{\underline A}  
\def\unb{\underline b}\def\unB{\underline B}  
\def\unc{\underline c}\def\unC{\underline C}  
\def\und{\underline d}\def\unD{\underline D}  
\def\une{\underline e}\def\unE{\underline E}  
\def\unf{\underline{\phantom{e}}\!\!\!\! f}\def\unF{\underline F}  
\def\unm{\underline m}\def\unM{\underline M}  
\def\unn{\underline n}\def\unN{\underline N}  
\def\unp{\underline{\phantom{a}}\!\!\! p}\def\unP{\underline P}  
\def\unq{\underline{\phantom{a}}\!\!\! q}  
\def\unQ{\underline{\phantom{A}}\!\!\!\! Q}  
\def\unH{\underline{H}}  
  
\def\As {{A \hspace{-6.4pt} \slash}\;}  
\def\bs {{b \hspace{-6.4pt} \slash}\;}  
\def\Ds {{D \hspace{-6.4pt} \slash}\;}
\def\Gts {{\Gt \hspace{-6.4pt} \slash}\;}
\def\ds {{\del \hspace{-6.4pt} \slash}\;}  
\def\ss {{\s \hspace{-6.4pt} \slash}\;}  
\def\ks {{ k \hspace{-6.4pt} \slash}\;}  
\def\ps {{p \hspace{-6.4pt} \slash}\;}   
\def\xs {{x \hspace{-6.4pt} \slash}\;}  
\def\pas {{{p_1} \hspace{-6.4pt} \slash}\;}  
\def\pbs {{{p_2} \hspace{-6.4pt} \slash}\;}   
\def\cFs {{{\cal F} \hspace{-6.4pt} \slash}\;}

\def\Ah{{\hat{A}}}  
\def\Dh{{\hat{D}}}
\def\Gh{{\hat{G}}}
\def\Fh{{\hat{F}}}
\def\Ih{{\hat{I}}} 
\def\Jh{{\hat{J}}} 
\def\Kh{{\hat{K}}}
\def\Lh{{\hat{L}}} 
\def\Ph{{\hat{P}}}
\def\Rh{{\hat{R}}}
\def\Vh{{\hat{V}}} 
\def\Xh{{\hat{X}}}
 
\def\ah{{\hat{\a}}}
\def\bh{{\hat{\b}}}
\def\gh{{\hat{\g}}}
\def\dh{{\hat{\d}}}
\def\rh{{\hat{\r}}}
\def\hh{\hat{h}}
\def\uh{\hat{u}}  
\def\xh{\hat{x}}  
\def\yh{\hat{y}}  
\def\ph{\hat{p}}  
\def\xih{\hat{\xi}}  
\def\chih{\hat{\chi}}  
\def\Psih{\hat{\Psi}}    
\def\phih{\hat{\phi}}

\def\psit{\tilde{\psi}}  
\def\Psit{\tilde{\Psi}}   
\def\Psibt{\tilde{\bar{Psi}}}  

\def\st{\tilde{\sigma}}  

\def\delt{\tilde{\delta}}
\def\Phit{\tilde{\Phi}}   
\def\Phitb{\overline{\tilde{Phi}}}  
\def\tht{\tilde{\th}}  
\def\lt{\tilde{\l}}
\def\chit{\tilde{\chi}}   
\def\phit{\tilde{\phi}} 

\def\At{\tilde{A}}
\def\Bt{\tilde{B}}
\def\Ct{\tilde{C}}
\def\Dt{\tilde{D}}
\def\Et{\tilde{E}}
\def\Ft{\tilde{F}}
\def\Gt{\tilde{G}}
\def\Ht{\tilde{H}}
\def\It{\tilde{I}}
\def\Jt{\tilde{J}}
\def\Qt{\tilde{Q}}  
\def\Rt{\tilde{R}}  
\def\Mt{\tilde{M }}  
\def\Nt{\tilde{N}}   
\def\St{\tilde{S}}
\def\Vt{\tilde{V}}
\def\Xt{\tilde{X}} 
\def\at{\tilde{a}}
\def\ct{\tilde{c}}
\def\dt{\tilde{d}}
\def\htt{\tilde{h}} 
\def\ft{\tilde{f}}
\def\gt{\tilde{g}}
\def\pt{\tilde{p}}  
\def\qt{\tilde{q}}  
\def\vt{\tilde{v}}  
\def\nt{\tilde{n}}  
\def\ut{\tilde{u}}  
\def\wt{\tilde{w}}  
\def\zt{\tilde{z}} 
\def\xt{\tilde{x}} 
\def\yt{\tilde{y}} 
\def\Psit{\tilde{\Psi}}
\def\vphit{\tilde{\varphi}}  

\def\eb{\bar{\epsilon}} 
\def\delb{\bar{\partial}}  
\def\thb{\bar{\theta}}
\def\mub{\bar{\mu}}
\def\lamb{\bar{\l}}
\def\psib{\bar{\psi}}
\def\sb{\bar{\sigma}}
\def\xib{\bar{\xi}}
\def\chib{\bar{\chi}}

\def\Psib{\bar{\Psi}}
\def\Phib{\bar{\Phi}}
\def\Lamb{\bar{\Lambda}}
\def\Sb{{\overline \Sigma}}
\def\cb{\bar{c}}
\def\hb{\bar{h}}
\def\qb{\bar{q}}
\def\wb{\bar{w}}
\def\ub{\bar{u}}
\def\zb{{\bar{z}}}
\def\Hb{\bar{H}}
\def\Qb{{\bar Q}}
\def\Omegab{\overline{\Omega}}
\def\ob{\overline{\omega}}

\def\Ab{{\overline A}} \def\Bb{{\overline B}} \def\Cb{{\overline C}}  
\def\Db{{\overline D}} \def\Eb{{\overline E}} \def\Fb{{\overline F}}  
\def\Gb{{\overline G}} 
\def\Ib{{\overline I}}  
\def\Jb{{\overline J}} \def\Kb{{\overline K}} \def\Lb{{\overline L}}  
\def\Mb{{\overline M}} \def\Nb{{\overline N}} \def\Ob{{\overline O}}  
\def\Pb{{\overline P}}  \def\Rb{{\overline R}}  
 \def\Tb{{\overline T}} \def\Ub{{\overline U}}  
\def\Vb{{\overline V}} \def\Wb{{\overline W}} \def\Xb{{\overline X}}  
\def\Yb{{\overline Y}} \def\Zb{{\overline Z}}  

\def\fb{{\overline f}}
\def\gb{{\overline g}}
\def\mb{{\overline m}}
\def\lb{{\overline l}}
\def\yb{{\overline y}}
  
\def\ldel{{\overleftarrow{\del}}}
\def\rdel{{\overrightarrow{\del}}}
\def\ldeldel{{\overleftarrow{\del^2}}}
\def\rdeldel{{\overrightarrow{\del^2}}}
\def\ldelb{{\overleftarrow{\bar{\del}}}}
\def\rdelb{{\overrightarrow{\bar{\del}}}}

\def\ba{{\bf a}} 
\def\bk{{\bf k}}  
\def\bl{{\bf l}}  
\def\bp{{\bf p}}  
\def\bq{{\bf q}}  
\def\br{{\bf r}}
\def\bt{{\bf t}}
\def\bu{{\bf u}}
\def\bv{{\bf v}}
\def\bx{{\bf x}}  
\def\by{{\bf y}}
\def\bA{{\bf A}} 
\def\bB{{\bf B}} 
\def\bR{{\bf R}}  
\def\bV{{\bf V}}  
 
\def\bz{{\boldsymbol{\zeta}}} 
 
\def\bone{{\bf 1}}  

\def\va{{\vec a}}
\def\vk{{\vec k}}
\def\vp{{\vec p}}
\def\vq{{\vec q}}
\def\vx{{\vec x}}
\def\vy{{\vec y}}
\def\vu{{\vec u}}
\def\vv{{\vec v}}
\def \vH{{\vec H}}
\def \vg{{\vec g}}

\def\vs{{\vec \sigma}}
\def\vtau{{\vec \tau}}

\newcommand{\ov}[1]{\overrightarrow{#1}}

\def\frA{\mathfrak{A}}
\def\frB{\mathfrak{B}}
\def\frC{\mathfrak{C}}
\def\frD{\mathfrak{D}}
\def\frE{\mathfrak{E}}
\def\frF{\mathfrak{F}}
\def\frG{\mathfrak{G}}
\def\frH{\mathfrak{H}}
\def\frM{\mathfrak{M}}
\def\frN{\mathfrak{N}}
\def\frR{\mathfrak{R}}
\def\frW{\mathfrak{W}}

\def\fra{\mathfrak{a}}
\def\frb{\mathfrak{b}}
\def\frf{\mathfrak{f}}
\def\frg{\mathfrak{g}}
\def\frh{\mathfrak{h}}
\def\frl{\mathfrak{l}}
\def\frs{\mathfrak{s}}
\def\fri{\mathfrak{i}}
\def\frj{\mathfrak{j}}

\def\ma{\mathfrak{a}}
\def\mg{\mathfrak{g}}
\def\mh{\mathfrak{h}}
\def\mR{\mathfrak{R}}
\def\mN{\mathfrak{N}}

\newcommand{\nn}{{\nonumber}}

\def\d{\delta}\def\D{\Delta}\def\ddt{\dot\delta}  
  
\def\pa{\partial} \def\del{\partial}  
\def\xx{\times}  
\def\uno{\mbox{1 \kern-.59em {\rm l}}}    
  
\def\trp{^{\top}}  
\def\inv{^{-1}}  
\def\dag{\dagger}  
\def\pr{^{\prime}}  
  
\def\rar{\rightarrow}  
\def\lar{\leftarrow}  
\def\lrar{\leftrightarrow}  
  
\newcommand{\0}{\,\!}      
\def\one{1\!\!1\,\,}  
\def\im{\imath}  
\def\jm{\jmath}  
  
\newcommand{\tr}{\mbox{tr}}  
\newcommand{\slsh}[1]{/ \!\!\!\! #1}  
  
\def\vac{|0\rangle}  
\def\lvac{\langle 0|}  
  
\def\hlf{\frac{1}{2}}  
\def\ove#1{\frac{1}{#1}}  
\newcommand{\hot}[1]{\frac{#1}{2}}

\def\Box{\square}  
\def\CC {\mathbb{C}}
\def\FF {\mathbb{F}}
\def\RR{\mathbb{R}}
\def\NN{\mathbb{N}}  
\def\ZZ{\mathbb{Z}}  
\def\bb#1{{\bf #1}}  
\def\bcomment#1{}  
\def\bfhat#1{{\bf \hat{#1}}}  
\def\VEV#1{\left\langle #1\right\rangle}  

\newcommand{\ex}[1]{{\rm e}^{#1}} \def\ii{{\rm i}}  

\newcommand{\lrbrk}[1]{\left(#1\right)}
\newcommand{\lrsbrk}[1]{\left[#1\right]}
\newcommand{\sfrac}[2]{{\textstyle\frac{#1}{#2}}}
 
\def\stw{{\sqrt{2}}}

\def\rf {{\rm f}}
\def\ri {{\rm i}}
\def\rj {{\rm j}}
\def\rn {{\rm n}}
\def\rk {{\rm k}}
\def\rl {{\rm l}}
\def\rs {{\scriptscriptstyle \rm S}}
\def\rt {{\scriptscriptstyle \rm T}}

\def\rQ {{\scriptscriptstyle \rm \cQ}}
\def\rR {{\scriptscriptstyle \rm \cR}}

\def\cQb{{\cal \Qb}}
\def\cRb{{\cal \Rb}}
\def\cWb{{\cal \Wb}}

\def\fd {{\rm N}}
\def\afd {{\overline{\rm N}}}

\def \II {I\hspace{-.1em}I\hspace{.1em}}
\def \IIA {\mbox{\II A\hspace{.2em}}}
\def \IIB {\mbox{\II B\hspace{.2em}}}
\def \gs {g^s}
\def \ls {\lambda^s}

\def \I {{\cal I}}
\def \qs {q\hspace{-.53em}/\hspace{.15em}}
\def \ks {k\hspace{-.53em}/\hspace{.15em}}
\def \YM {{\mbox{\tiny YM}}}
\def \gym {g_{\YM}}

\def \Lc {\L_c}
\def\IR{\relax{\rm I\kern-.18em R}}
\def \id {{\bf 1}}

\def\cci{\ell}
\def\ccj{\ell'}

\def\bbq{\pmb{q}}

\begin{document}
\begin{titlepage}
\begin{flushright}
\hfill{NCTS-CMT/1505}
 \end{flushright}
\hfill 

 \begin{center}
 
{\Large \bf Weyl Semimetal and \\  
Nonassociative Nambu Geometry
}\\[10mm] 

{\bf Chong-Sun Chu${\,}$}

{\itshape Physics Division, National Center for Theoretical Sciences, \\
 National Tsing-Hua University, Hsinchu, 30013, Taiwan}\\[1mm]
{\itshape and}\\
{\itshape Department of Physics, National Tsing-Hua
  University,  Hsinchu 30013, Taiwan}\\

\end{center}

\date{\today}

\begin{abstract}
Topological materials are characterized by an electronic  
band structure with
nontrivial topological properties. In this paper 
we introduce a 
basis of operators for the linear space of operators 
spanned by charge-neutral fermion bilinears. These band projected density 
operators are constructed using directly the eigenfunctions of the
electronic 
energy band structure and there is no need to assume a flat Berry
curature. As a result,  our set of operators has a wider range of validity and
is sensitive to physical phenomena which are not detectable in the flat
curvature limit. In particular, we show that the 
Berry monopole configuration of Weyl semimetal give rises to a
nonvanishing Jacobiator for these band projected density operators,
implying the emergence of nonassociativity at the location  of the
Weyl nodes. The resulting nonassociativity observes the fundamental identity,
the defining property of the Nambu bracket and so one may call this a
nonassociative Nambu geometry. We also derive the corresponding 
uncertainty principle. 

\end{abstract}

\end{titlepage}
\newpage


\section{Introduction}

Topological material is one of the recent hot topics of research in
condensed matter physics. These are novel 
electronic states of matters with properties
supported by topology. Topological material
has  not only overturned the traditional Landau's paradigm on
the classification of condensed matter states, it has also 
many remarkable properties that are not only interesting
theoretically, but has also  important potential practical
applications, such as topological quantum computing \cite{bernevig}. 

Topological insulator (TI) is the simplest form of   topological
material.
Different from an ordinary insulator, TI is
conducting at the boundary due to the existence of gapless chiral
states at the surface, which in turn is  a result of the
nontrivial momentum space topology of the bulk band structure.
As long as the band gap is not closed, the topological ground state is
robust against small 
perturbations and  the surface states are
protected.
The Chern Insulator (CI) is
the most primitive example of a topological insulator. As  two
dimensional system, it has a band structure which is characterized by
the first Chern number of the Berry connection associated with the
Bloch waves, and exhibits a quantized Hall conductance 
even without an external magnetic field \cite{haldane88}, 
generalizing the original Integer Quantum Hall Effect (IQHE) 
\cite{KDP80}
for the filled Landau levels \cite{thouless82}. 
In the presence of time reversal symmetry (TRS),
the first Chern numbers and the Hall conductivity must vanish. However
the spin-orbit interaction allows a different topological class of
insulating band structures, giving rises to the 
$\ZZ_2$ topological insulator in two and three dimensions 
\cite{kane2005,fu2007,moore2007,roy2009}, leading to successful
prediction and experimental observation of these phases of materials 
\cite{e1,e2,e3,e4,e5,e6}.

A slightly more complicated and also much more interesting topological
material is the Fractional Chern Insulator (FCI), where 
the topological bands are partially filled. The best known example is
the the Fractional Quantum Hall Effect (FQHE) \cite{tsui}, a strongly
correlated phenomena which arises
from the huge degeneracy within the Landau levels for particles in a
magnetic field. Here a very useful observable is the density
operator projected to the lowest Landau level. These operators were 
introduced originally by  Girvin, Macdonald and Platzman (GMP)
\cite{GMP85}, and were found to
obey  a $W_\infty$ Lie algebra \cite{GMP86}, which reflects
the area preserving incompressible nature 
of the Laughlin wavefunctions of the FQHE \cite{laughlin}.
Inspired
by the success of the algebraic approach 
of GMP to the problem of IQHE and
FQHE, Haldane 
proposed recently 
a geometrical description of the FQHE based on the algebra of these
projected density operators \cite{haldane2011}. 
More recently,  
a specific noncommutative geometry for the band-projected density
operators   was identified by  Parameswaran, Roy and
Sondhi \cite{PRS1}
for general two-dimensional CI and FCI. 
See also \cite{a1,a2,gmp1,gmp2} for related algebraic 
approaches.   A reformulation of the
Hamiltonian theory of the FQHE was introduced in
\cite{mur-fermion, gmp3}. Our work is motivated by these studies.

The noncommutative geometry identified by \cite{PRS1}
generalizes the well known noncommutative
geometry obeyed by the guiding center of the Landau level electron
$[X,Y] = i l_B^2$, 
where $l_B = \sqrt{\hbar/eB}$ is the magnetic length. 
In the limit of long wavelength and flat Berry curvature, 
it coincides with the noncommutative
geometry of GMP.  The noncommutative geometry is expected to be useful
since in the limit the band
structure admit a large
band gap compared to the interaction strength, which is relevant for
the studies of the FQHE, it makes sense
to project the problem to the lowest filled band where the major
effects of the interaction take place. Due to the universality of the
identified noncommutative geometry,  Parameswaran, Roy and
Sondhi also proposed  that the band noncommutative geometry
could be  useful to the study of the FQHE in FCI in a similar way.
The noncommutative geometry of topological insulator was then 
further studied  in the three and the higher-dimensional case
\cite{ryu,d-alg,h2,h3,h4,h5,h6}. 
In particular, an interesting 3-bracket utilizing the completely
antisymmetrizer has been
constructed  for the differentiated projected density operators in
3-dimensions in \cite{ryu}.

We  remark that 
the above mentioned noncommutative geometries were obtained as
a property
of the electronic band. They are purely kinematical  and independent
of interaction. Furthermore it does
not matter whether the material is topological or not, the
noncommutative geometry  takes on the
same universal form with the dependence
on the materials enters only through the
band's Berry connection. This motivates us to ask if there exists a
more refined characterization of topological materials 
that is intrinsic to the nontrivial topology. We find in this paper 
that
Weyl semimetal is characterized by a nonassociative geometry
\eq{rrr-3d} of 
the projected density operators.

Weyl semimetal (WSM) is an interesting class of topological
materials that lives in 3 dimensions.
Unlike topological insulator, 
the valence band and the conduction band of WSM touches
at isolated points in the momentum space.
The nodal point is called a Weyl node since 
the nearby band structure is described by the
Hamiltonian of  a massless Weyl fermion, and, depending on the
chirality of the node, the associated Berry connection
describes a monopole or anti-monopole in the 3-dimensional momentum
space. Weyl nodes are topological objects. 
The only way for the Weyl nodes to
disappear is to annihilate them in pair. In fact, 
the theorem of Nielsen and Ninomiya states that Weyl nodes
always come in pairs of opposite chirality \cite{nielsen}. Therefore to
obtain a stable WSM, its Weyl nodes need to stay separated 
in the momentum space in order to prevent them from annihilation.  This 
can be achieved when either TR or inversion symmetry is broken. 
In the presence of both TR and inversion symmetry, the
touched band must be doubly degenerated and one obtain a Dirac fermion
spectrum near the nodal points, resulting in the so called Dirac
semimetal (DSM).  

As a result of the topologically nontrivial band structure, like any
other topological materials, 
WSM is also endowed with topologically protected surface
states, the Fermi arcs \cite{wan}. 
One particularly illuminating way \cite{burkov2011} 
to understand the origin of the
Fermi arcs is to consider a slicing of the Brillouin Zone (BZ).  
Each momentum space slice of the WSM which does not contain the Weyl
nodes are Chern insulators whose Chern number changes by $\pm 1$ as one
sweeps past a Weyl node. Thus if the slices in between the nodes have 
a unit Chern number, i.e. these slices are nontrivial CI, 
then the Fermi arcs
are simply the edge states of these CI's. The simple 
connection between Weyl semimetal and Chern insulator illustrated in this
picture  also
suggest that some of what we know about two dimensional CI could 
be and should be 
generalized to the 3 dimensional case. This is another motivation
of this paper.

The planning of this paper is as follows.  In section 2.1, we first 
review the noncommutative geometry for band projected density
operators. We also generalize the result to 
more general band projected operators. In section 2.2, we compute the
Jacobiator for the band projected operators and find that it is
nonzero in the presence of monopole. This is the case as  in the Weyl
semimetal. We also explain the origin of the
nonassociativity. In section 2.3, we show that the obtained Jacobiator 
satisfies the fundamental identity, the defining property of the Nambu
bracket \cite{nambu73}. As a result, we call this the
{\it nonassociative Nambu  geometry}. In section 3,
we consider the uncertainty principle of the nonvanishing 
Jacobiator. Just as a nonvanishing commutator leads to the Heisenberg
uncertainty principle that involves a 
product of the uncertainties
of the operators appearing in the commutator, 
we show that a nonvanishing Jacobiator leads
to similar uncertainty principles that involves a  
product of the uncertainties
of the operators appearing in the Jacobiator.

\section{Quantum Geometry of Topological Band}

In the band theory of crystal structure, motion of a single electron
can be approximated by treating the whole lattice  of ions and other
electrons as a static background. We consider an insulator with $N$
bands. The single particle
Hamiltonian takes the form
\be
H_0 = \sum_{a,b,\bk} c^\dag_{\bk, a} h_{ab} (\bk) c_{\bk, b},
\ee 
where 
$a,b = 1 \cdots, N$ label the states in the unit cell,
$\bk = (k_1, \cdots, k_D)$ is the single particle momentum
restricted to the first Brillouin Zone 
and $D$ is the (spatial) dimension
of the material.
The Hamiltonian can be diagonalized straightforwardly by considering
the eigenvalue problem 
\be
\sum_b h_{ab} (\bk) u_b^\a (\bk) = E_\a(\bk) u_a^\a (\bk),
\ee
where $\a = 1, \cdots, N$, labels the band energy 
$E_\a(\bk)$.
Adopting the orthonormality condition for the eigenfunctions
\be
\sum_a u^\a_a(\bk){}^* u_a^{\b} (\bk) = \d^{a\b},
\ee
then the  orbital creation operator
\be
\g^\a_\bk{}^\dag : =\sum_a u_a^\a(\bk) c^\dag_{\bk, a}
\ee
obeys
\be
[\g^\a_\bk{}^\dag, \g_{\bq}^\b]_+ = \d_{\bk,\bq} \d^{\a\b},
\ee
and the Hamiltonian can be written in the diagonalized form
\be \label{H0-diag}
H_0 = \sum_{\bk, \a} E_\a(\bk) \g^\a_\bk{}^\dag \g^\a_\bk
\ee
with the eigenstates
\be \label{kets}
\ket{\bk, \a} = \g_{\bk,\a}^\dag \ket{0}.
\ee

Despite the simple appearance of \eq{H0-diag}, 
the information encoded in the eigenfunctions
$u_b^\a(\bk)$ is not lost. 
One of the remarkable feature of the energy band structure is that it 
is naturally equipped 
with a Berry connection. For a given band $\a$, the Berry
connection is defined by
\be
A^\a_j(\bk) = i \sum_{b=1}^N u_b^\a(\bk){}^* \frac{\del}{\del
  {k_j}} u_b^\a(\bk), \quad j = 1, \cdots, D. 
\ee
The definition can be straightforwardly generalized to a nonabelian 
Berry
connection involving an arbitrary number of bands. 
In the standard application, the Berry connection is kinematical.
Taking into account of the fluctuation of the crystal ions, one may
wonder whether a kinetic term would be induced as in induced gravity 
\cite{sak}. 

\subsection{Noncommutative geometry for band projected operators}


Making use of the projection operator 
$P_\a = \sum_\bk \ket{\bk, \a} \bra{\bk,\a}$, 
one can project the density operator $\rho_\bq = 
e^{ i\bq \cdot \br}$ onto the band $\a$. In momentum space, the projected
density operator takes the form \cite{PRS1}
\be
\rh_{\bq, \a} := P_\a \r_\bq P_\a = \sum_{\bk, b} 
u_b^{\a*}(\bk +\frac{\bq}{2}) u_b^\a(\bk -\frac{\bq}{2}) \times 
\g_{\bk+\frac{\bq}{2} }^{\a \; \dag} \g_{\bk -\frac{\bq}{2} }^\a. 
\ee
It is 
\be
\rh_{\bq, \a}^\dag = \rh_{- \bq, \a}.
\ee
In the following we will focus on a single  band and so we will
skip the subscript $\a$ and simply write $\rh_{\bq, \a}$ as   $\rh_{\bq}$.
In the paper \cite{PRS1}, it was found that the density operator at
different momentum obeys,  in the leading order
of small momentum, the commutator relation:
\be \label{rr}
[\rh_{\bq_1}, \rh_{\bq_2}] = i q_1^i q_2^j \sum_{\bk,b} F_{ij}(\bk) 
u_b^{\a*}(\bk+ \frac{\bbq}{2}) u_b^\a(\bk - \frac{\bbq}{2} ) \times 
\g_{\bk + \frac{\bbq}{2} }^{\a \;\dag} \g_{\bk - \frac{\bbq}{2}}^\a, 
\ee
where
\be
F_{ij}(\bk) : = \del_i A_j(\bk) - \del_j A_i(\bk) 
\ee
is the curvature of the Berry connection, and 
$\bbq := \sum_n \bq_n = \bq_1 +\bq_2$ is the sum of the momentum of the
operators on the RHS of \eq{rr}. 
The noncommutative relation \eq{rr} is interesting. It is an universal
property of band insulator.
It takes the same form even when the material is non-topological with
vanishing 
Chern numbers. In the literature sometimes it is considered
the situation of having a Berry curvature slowly varying over the BZ. In this
case, \eq{rr} takes the
approximate form
\be \label{rr-approx}
[\rh_{\bq_1}, \rh_{\bq_2}] \approx i q_1^i q_2^j \bar{F}_{ij}\;  
\rh_{\bq_1+ \bq_2},
\ee
where $\bar{F}_{ij}$ is the mean value of the Berry curvature over the
BZ. The result \eq{rr-approx} is now topological: it is non-trivial only
when the 
Chern number is nonvanishing. 
In this paper,
we will be
interested in the general behaviour of the Berry connection without making
any assumption that it is slowly varying. In particular, our main
results, \eq{FFF}, \eq{FFF-3d}  about the Jacobiator 
are non-trivial only in the presence of monopoles and non-trivial topology.
This is in contrast with the noncommutative geometry \eq{rr} which can
be non-trivial even in the absence of non-trivial topology. 

It is instructive to generalize the result \eq{rr} for the commutator
of two projected density operators. In general for any arbitrary function
$f(\bk)$  defined
on the BZ and momentum $\bq$, one can introduce the band projected operator
\be \label{cO}
\cO(f,\bq) = \sum_{\bk, b} f(\bk) 
u_b^{\a *}(\bk + \hot{\bq} ) u_b^\a(\bk -\hot{\bq}) 
\times  \g_{\bk+\hot{\bq}}^{\a\;\dag} \g_{\bk -\hot{\bq}}^\a
\ee
in association with the band $\a$.
As explained  above, we will ignore the band index.
Note that since  $\cO$ depends linearly on its first
argument, it is obvious that 
\be
\cO(f,\bq) + \cO(g,\bq) = \cO(f+g, \bq). 
\ee
It is obvious that the set of operators \eq{cO}
form an over-complete basis for the linear space of operators 
spanned by charge-neutral fermion bilinears.
The operators \eq{cO}
provide a generalization of the projected density
operator and can be used to
describe more complicated interaction other than the density-density
type. 
It is natural to ask if the set $\cA$ of operators $\cO$ 
for all functions $f$ defined on the BZ form a Lie algebra; and if so,
of what kind.

Therefore, let us consider the commutator. 
It is clear from the definition \eq{cO} of the operator that the
commutator of these operators gives something that is bi-linear in the
orbital creation operator
\be \label{FF-general}
[\cO(f_1, \bq_1), \cO(f_2, \bq_2) ] = \sum_{\bk, b, c} K_{bc}(\bk,
\bq_1,\bq_2) \times  \g_{\bk_+}^{\a \; \dag} \g_{\bk_-}^\a,
\ee
where $\bk_\pm : = \bk \pm \hot{\bbq}$,\; $\bbq:= \bq_1+ \bq_2$ 
and $K$ is some 
kernel depending on $b,c$ and the momentum $\bq_1, \bq_2$ and
$\bk$. It  is 
given by
\begin{multline} \label{cK}
K_{bc}(\bk, \bq_1,\bq_2) = 
f_1(\bk-\hot{\bq_2}) f_2(\bk+ \hot{\bq_1}) u_b^*(\bk_+ -\bq_2)
u_b(\bk_-) u_c^*(\bk_+)u_c(\bk_-+ \bq_1) \\
- f_1(\bk+\hot{\bq_2}) f_2(\bk- \hot{\bq_1}) u_b^* (\bk_+) 
u_b(\bk_- +\bq_2) u_c^*(\bk_+ -\bq_1)u_c(\bk_-).
\end{multline}
In general $K_{bc}$ is not diagonal and thus the commutator of
$\cO$'s does not close back to $\cO$. However, in the
approximation of small momentum $\bq_1, \bq_2$, one can expand $K$. It
turns out that the leading term in the approximation of small momentum 
is diagonal in $b,c$ and takes the simple form
\begin{multline}
K_{bc}(\bk, \bq_1,\bq_2) \\
= \d_{bc} \; \Big(
i q_1^i q_2^j  (F_{ij} f_1 f_2)(\bk)
+ q_1^i (f_1 \del_i f_2 )(\bk) - q_2^i (f_2 \del_i f_1 )(\bk) 
\Big)
u_b^{\a*}(\bk + \hot{\bbq} ) u_b^\a(\bk -\hot{\bbq})  +\cdots,
\end{multline}
where $\cdots$ 
denotes terms that are higher order in the momentum $\bq_1$ or $\bq_2$.
As a result, in the leading order of small momentum, 
the operators $\cO$ obeys the operator relation \eq{FF},
\be \label{FF}
[\cO(f_1, \bq_1), \cO(f_2, \bq_2) ] =
\cO(\{ (f_1,\bq_1), (f_2,\bq_2) \},\bq_1 + \bq_2),
\ee
where the 2-bracket $\{,\}$ is defined by
\be \label{ff}
\{(f_1,\bq_1), (f_2,\bq_2) \}  
:= ( i q_1^i q_2^j  (F_{ij} f_1 f_2) 
+ q_1^i (f_1 \del_i f_2 ) - q_2^i (f_2 \del_i f_1 ) , \bq_1+ \bq_2).
\ee
The bracket $\{,\}$ is defined for any pair of objects $(f_1,\bq_1), (f_2,\bq_2) $
where $f_n$'s are functions defined on the BZ and $\bq_n$'s are
momentum restricted to  the BZ.
Note that  the momentum  $\bq_1, \bq_2$ enter
linearly in the 2-bracket $\{, \}$. 
Effectively, we have shown that, 
in the leading order of small momentum, the set $\cA$ 
can be endorsed with a commutator, with respect to which 
it becomes a closed algebra. Furthermore, the commutation relation of
$\cO$ induces a  2-bracket $\{ , \}$  
on the pair of objects $(f, \bq)$.

We remark that a different set of fermion bilinear operators has been
introduced before 
\cite{gmp1,gmp2, gmp3}. They form a a Lie algebra, and, for even
dimensions, constitutes a complete basis \cite{gmp2}. 
In contrast to these operators, which were 
suitably defined in the limit of a flat Berry curvature,
our operators \eq{cO} are constructed directly using the
eigenfunctions of the electronic energy band and there is no need to 
assume a flat Berry curvature. 
Because of this, our set of operators has a wider range of validity and
is sensitive to physical phenomena which are not detectable in the flat
curvature limit. In particular, our set of operators could provide
support to a kind of nonassociative geometry which occurs in the
presence of monopoles. This is another main result of this paper. 

\subsection{Non-associative geometry and monopoles}

To decide whether it is a Lie algebra, we need to check the
Jacobi identity. After a long but straightforward computation, we find
\be \label{fff}
 \{ \{ (f_1, \bq_1), (f_2, \bq_2) \}, (f_3, \bq_3) \} + \mbox{cyclic}
= \big(- i f_1 f_2 f_3 \sum_{i,j,k =1}^D q_1^i q_2^j q_3^k (\del_i F_{jk} 
+ \mbox{cyclic}), \bbq\big) .
\ee
Here
$ \bbq:= \sum_n \bq_n =\bq_1+\bq_2 + \bq_3 $ is the sum of momentum of
the operators on the LHS of the equation \eq{fff}. 
We note that in principle the left hand side of \eq{fff} contains
terms of the form $fg q\del_i h$ with one derivative acting on $f,g$
or $h$, and  terms of the form $f_1 f_2 \del_i \del_j f_3$ and 
$f_1 \del_i f_2 \del_j f_3$  with two derivatives acting on the $f_n$'s. 
However all these terms get canceled with each other and in
the end only
the non-derivative term $ f_1 f_2 f_3$ is left over. Hence the result
\eq{fff} is exact and  there is no need to make any assumption of
small momentum.

It follows from \eq{fff} that
the Jacobiator for the operators $\cO$ takes the simple form,
\be \label{FFF}
[\cO(f_1, \bq_1), \cO(f_2, \bq_2),\cO(f_3,\bq_3) ]
=
- i f_1 f_2 f_3 \sum_{i,j,k =1}^D q_1^i q_2^j q_3^k \; \cO \Big(
\del_i F_{jk}  + \mbox{cyclic} , \bbq \Big) \;
\ee
in the leading order of small momentum.
Here the Jacobiator
for any three operators $A,B,C$ is defined as
\be
[A,B,C] := [[A,B],C] + [[B,C],A] + [[C,A],B]. 
\ee
For 3 dimensions, we have
\be \label{fff-3d}
 \{ \{ (f_1, \bq_1), (f_2, \bq_2) \}, (f_3, \bq_3) \} + \mbox{cyclic}
 =
(- i f_1 f_2 f_3 \; (\bq_1 \times \bq_2) \cdot \bq_3
\; \nabla \cdot {\bf B}, \; \bbq)
\ee
and 
\bea \label{FFF-3d}
[\cO(f_1, \bq_1), \cO(f_2, \bq_2),\cO(f_3,\bq_3)]  = 
- i f_1 f_2 f_3\; (\bq_1 \times \bq_2) \cdot \bq_3 \; \cO \Big(
\nabla \cdot {\bf B} ,\; \bbq \Big)
\eea
in the leading order of small momentum.
We note that the right hand side of \eq{fff}, \eq{FFF}, \eq{fff-3d}
and  \eq{FFF-3d} are zero for smooth
configurations of the Berry connection since the Berry curvature $F_{ij}$
is a total derivative. In this case the 2-bracket $\{, \}$ defines a Lie
bracket on $\cA$ and the operator product between the operators $\cO$ is
associative. 
However in the presence of a monopole, which is characteristic of
topological insulator,  the Jacobi identity \eq{fff} is
violated and the 2-bracket $\{, \}$ does not define a Lie bracket. 
Correspondingly the
operator algebra $\cA$   is nonassociative.

The projected density corresponds to the simplest case of a constant 
function $f=1$,
\be
\rh_\bq = \cO(1,\bq)
\ee
and the result \eq{rr} follows immediately 
from the 2-bracket \eq{ff} that
\be
\{(1,\bq_1), (1,\bq_2) \} =  i q_1^i q_2^j  F_{ij} .
\ee
As for the Jacobiator, we have for 3-dimensions the result 
\be \label{rrr-3d}
[\rh_{\bq_1}, \rh_{\bq_2}, \rh_{\bq_3}] 
= -i (\bq_1 \times \bq_2) \cdot \bq_3 \sum_{\bk, b} 
\nabla \cdot {\bf B} (\bk) 
u_b^{\a*}(\bk + \frac{\bbq}{2}) 
u_b^\a(\bk - \frac{\bbq}{2} ) \times
\g_{\bk +  \frac{\bbq}{2}}^{\a\,\dag} 
\g_{\bk -  \frac{\bbq}{2}}^\a.
\ee
We comment that in contrast to the commutation relation  \eq{rr} which
takes the same form universal to all energy band,
the violation of nonassociativity spotted by the Jacobiator \eq{rrr-3d}
is an intrinsic characterization of topological Weyl semimetal. 
 
The emergence of nonassociativity is not a new phenomena in physics. 
As far as we know, the Jacobiator first appeared in the literature of
particle physics and quantum field theory in the computation of 
the space components of current in the quark model \cite{q1,q2}.
Later, a proper understanding of
the Jacobiator in terms of the 3-cocycle of an associated
(nonassociative) group transformation was developed in
\cite{j1,j2,j3,j4}. Moreover, as an example, 
the quantization  of a charged particle
in a magnetic monopole is shown to give rise to a nonvanishing
Jacobiator. Let us recall briefly this result. 
Consider a charged
particle in the presence of an external magnetic field $\bB$ in 3
dimensions.  A
gauge invariant canonical momentum does not exist. Instead  the
velocity operator $v_i = (p_i + e A_i)/m$ is gauge invariant. 
The velocities do not commute
\be \label{vv}
[v_i, v_j] = i \frac{e \hbar}{m^2}\e_{ijk} B_k,
\ee
and have the Jacobiator
\be \label{vvv}
[v_3, [v_1,v_2]] + \mbox{(123) cyclic} = \frac{e \hbar^2}{m^3} \nabla
\cdot \bB,
\ee
which is nonvanishing in the presence of magnetic monopole, $ \nabla
\cdot \bB = 4 \pi g \neq 0$.
The presence of nonassociativity in the operator algebra means that the
velocity operators are not globally defined. This is because the
vector potentials are not 
globally defined  in the presence of monopole; in which
case we can either use a singular description involving a Dirac
string, or equivalently, use the Wu-Yang description which 
employs two patches of
potential related by a gauge transformation.  At the level of gauge
bundle, the monopole charge $g$ must satisfy the Dirac
quantization condition $e g = \frac{\hbar}{2} \ZZ$. The same condition
guarantees that translations commute and a   proper quantum mechanical
formalism exists \cite{j1,j2,j3,j4}. 
 
Our consideration and analysis has in fact been motivated by the
knowledge of this simple
system. In particular the striking similarity of \eq{rr} with
\eq{vv} has leaded us to suspect that the Jacobiator of the projected
density operators will be proportional to $\nabla \cdot \bB$ and this
is indeed the case as we obtained in \eq{rrr-3d}. In our case,
however, the breakdown of the associativity is more basic. 
It is due to the
merging of the bands at the Weyl node, which results in a 
change of the degeneracy of the energy levels. Recall how the
projected density operator acts on the states \eq{kets} 
\be \label{reps}
\rh_{\bq} \ket{\bk, \a} = e^{i \int_\bk^{\bk+ \bq} d\bk' \cdot
\bA (\bk')} \ket{\bk+\bq, \a}
\ee
in the long wavelength limit. The occurrence of degeneracy leads to a
Berry monopole configuration \cite{berry-wik84}. 
This in principle is bad 
for \eq{reps} 
since the corresponding Berry connection contains a Dirac string
singularity. 
However it is well known that  if the Dirac quantization 
condition is satisfied, then the Dirac string singularity
becomes a gauge artifact, and hence the representation \eq{reps} is
well defined. Nevertheless the presence of the monopole is physical 
and we find that its singular nature is  felt through certain 
successive  action, the Jacobiator,  
of the band projected operators.


\subsection{Fundamental identity and Nambu geometry}

Next we would like to characterize the kind of 
nonassociative geometry \eq{FFF-3d} we are having here. 
In general, given an algebra with a binary product $\circ$, one can 
introduce the associator defined by
\be
\d (A,B,C) := (A \circ B) \circ C - A \circ (B \circ C)
\ee
to characterize  the nonassociativity of the algebra.
The associator is related to the Jacobiator as
\be
[A,B,C] := \d(A,B,C) + \d(B,C,A) + \d(C,A,B) - 
\d(B,A,C) - \d(C,B,A) - \d(A,C,B). 
\ee 
As we have just demonstrated, the algebra $\cA$ of operators $\cO$ is
non-associative in general. 
It is interesting to characterize the type of
the non-associative geometry. 
A first guess which come to
the mind is the Malcev algebra that has also appeared in some studies
in string theory \cite{minic}. 
In general, a Malcev algebra $\cA$ is an 
algebra equipped
with an antisymmetric product $\circ$, 
\be
x_1 \circ x_2 = - x_2 \circ x_1
\ee
in which the Malcev identity 
\be
(x_1 \circ x_2) \circ (x_1 \circ x_3) = (((x_1 \circ x_2) \circ x_3) 
\circ x_1) +
 (((x_2 \circ x_3) \circ x_1) \circ x_1) 
+  (((x_3 \circ x_1) \circ x_1) \circ x_2)
\ee
is satisfied for any $x_1, x_2, x_3 \in \cA$. 

In our case we can define an
antisymmetric product from the commutator
\be
x_1 \circ x_2 := [x_1,x_2].
\ee
Then, in terms of the Jacobiator, the Malcev identity reads
\be
[[x_1,x_2,x_3],x_1] = [x_1,x_2,[x_3,x_1]].
\ee
Taking $x_n = \cO(f_n, \bq_n)$, 
the operator relation is translated to the following 
statement on the 2-bracket
\be \label{jacobi-f}
\{ \{(f_1, \bq_1), (f_2, \bq_2), (f_3, \bq_3) \},  (f_1, \bq_1)\} 
= \{(f_1, \bq_1), (f_2, \bq_2), \{ (f_3, \bq_3), (f_1, \bq_1)\}\} ,
\ee
where
\be
\{(f_1, \bq_1), (f_2, \bq_2), (f_3, \bq_3)\} := 
\{ \{(f_1, \bq_1), (f_2, \bq_2)\}, (f_3, \bq_3)\} + \mbox{cyclic}
\ee
is the Jacobiator for the 2-bracket $\{ , \}$.
It is easy to check that \eq{jacobi-f} is not satisfied. Hence
the nonassociativity occurring in the topological insulator is not of
the Malcev type. 
We remark that the closest we can get for a
meaningful 
relation involving a Jacobiator and a 2-bracket
$\{,\}$ is
\begin{multline}
\{ \{(f_1, \bq_1), (f_2, \bq_2), (f_3, \bq_3)\},  (f_1, \bq_1)\} \\
= \{(f_1, \bq_1),  \{ (f_2, \bq_2), (f_1, \bq_1)\}, (f_3, \bq_3) \} +
\{(f_1, \bq_1), (f_2, \bq_2), \{ (f_3, \bq_3), (f_1, \bq_1)\} \} \\
+ (i f_1^2 f_2 f_3 (\bq_1 \times \bq_2)\cdot \bq_3 \; 
(\bq_1 \cdot \nabla) (\nabla \cdot {\bf B}), 2 \bq_1+ \bq_2+\bq_3).
\end{multline}
There is however 
not a simple type of nonassociativity that one can associate  this with. 

What about a relation involving two Jacobiators?
In the literature, given an algebra $\cA$ with a multilinear and
completely antisymmetric 3-bracket 
$\{\cdot , \cdot, \cdot  \}: \cA\otimes \cA \otimes \cA \to \cA $, 
the 3-bracket is
said to satisfy the {\it fundamental identity} if the bracket of
bracket satisfies the following relation
\be \label{3p}
\{\{x_1,x_2,x_3\},x_4,x_5\} 
= \{\{x_1,x_4,x_5\},x_2,x_3\} 
+ \{x_1,\{x_2, x_4, x_5\}, x_3\}+ \{x_1, x_2,\{x_3, x_4, x_5\}\}, 
\ee
for arbitrary $x_1, \cdots, x_5 \in \cA$. 
The fundamental identity is
a natural generalization of the {\it Jacobi identity} 
\be \label{2p}
\{\{x_1,x_2\},x_3\} = \{\{x_1,x_3\},x_2\} + \{x_1, \{x_2,x_3\}\}
\ee
for antisymmetric 2-bracket 
$\{\cdot , \cdot \}: \cA \otimes \cA \to \cA $.
The fundamental identity is an important consistency condition which
allow for the introduction of a symmetry transformation of the algebra
Just as an antisymmetric 2-bracket which satisfies the Jacobi identity
\eq{2p}  can be used to define a homeomorphism of the algebra
\be
\d X := \{ a,X \}, \quad a,X \in \cA,
\ee 
which 
acts as a derivation on the 2-bracket
\be
\d \{ X,Y \} = \{ \d X, Y\} + \{ X, \d Y\} , 
\ee
a 3-bracket  which satisfies the  the fundamental identity \eq{3p} 
can be used to
generate a homeomorphism of the algebra 
\be
\d X := \{ a,b, X \}, \quad a,X \in \cA.
\ee
This acts as a derivation on the 3-bracket
\be
\d \{ X,Y,Z \} = \{ \d X, Y, Z\} + \{ X, \d Y, Z\} + \{X, Y, \d Z \}
\ee
and can thus be considered as a symmetry transformation of the
algebra.

The simplest example
of a 3-bracket which satisfies the fundamental identity is the
canonical Nambu bracket $\{f,g,h \} := \e^{ijk} \del_i f \del_j g
\del_k h $ defined for any functions $f, g, h$ over a 3 dimensional
manifold \cite{nambu73}. In general a Nambu bracket is a completely
antisymmetric 3-bracket which 
satisfies the fundamental identity \eq{3p}.
It is straightforward to check that the Jacobiator \eq{fff-3d}
for any three  $x_i = (f_i,\bq_i), i =1,2,3$, 
indeed satisfies the fundamental  identity \eq{3p}.
As a result, the Jacobiator \eq{FFF-3d} also satisfies the fundamental
identity
\begin{multline} \label{3p-FFF}
[[\cO_1,\cO_2,\cO_3],\cO_4,\cO_5] \\
= [[\cO_1,\cO_4,\cO_5],\cO_2,\cO_3] 
+ [\cO_1,[\cO_2, \cO_4, \cO_5], \cO_3]+ 
[\cO_1, \cO_2,[\cO_3, \cO_4, \cO_5]],
\end{multline}
where $\cO_n$ denotes the operator $\cO(f_n,\bq_n)$.
Hence our nonassociative geometry, characterized by the Jacobiator
\eq{FFF-3d}, is a Nambu bracket and  we call this a nonassociative 
Nambu geometry.

We remark that the notion of Nambu bracket as one
characterized by the fundamental identity was originally introduced by
Nambu in his formulation of a generalized mechanics \cite{nambu73}. 
In the
same paper Nambu also considered the quantization problem of the
classical Nambu mechanics and considered the completely
antisymmetrizer as a
candidate for a quantization of the canonical Nambu bracket. This 
3-bracket was considered in \cite{ryu}. 
We also remark that in some applications, e.g. 
in the generalization of the Hamiltonian mechanics known as the 
Nambu mechanics \cite{tak}, 
it is useful to introduce the notion of a 
Nambu-Poisson bracket, which 
is defined for an algebra with a binary product
$\circ$ to be a completely antisymmetric 3-bracket which satisfies 
in addition to the fundamental identity also  the
derivation rule:
\be
\{ x_1, x_2 , y_1 \circ y_2 \} = y_1 \circ \{x_1,x_2,y_2\} +
\{x_1,x_2,y_1\}\circ y_2.
\ee
This is not what we considered here.

\section{Nonassociative Uncertainty Relation}

For a noncommutative geometry defined by a commutation relation
\be
[X,Y] = i \th,
\ee
it is easy to derive an associated uncertainty relation constraining
the quantum fluctuation of the operators.
If $X,Y$ are Hermitian,  the noncommutative 
uncertainty relation takes the form 
\be\label{ur-2}
\d X \; \d Y \geq \frac{1}{2} | \langle [X,Y] \rangle |.
\ee
This gives a  constraint on the root-mean-square deviation 
of the operators 
\be
\d X := \sqrt{\langle  (\D X)^2 \rangle}, \quad 
\D X:= X - \langle X \rangle,
\ee
in terms of  the
expectation value of the commutator of $X, Y$. 
The uncertainty relation
\eq{ur-2} can be easily generalized to the case of non-Hermitian
operators. However the generalization is not unique. For example one
can write down the following uncertainty relation
\be \label{ur-g1}
\frac{1}{2}(
\d X \; \d' Y + \d' X \; \d Y)  \geq
\frac{1}{2} 
\Big| \big \langle [X, Y] \big \rangle \Big|,
\ee
or the more symmetrical form
\be \label{ur-g2}
\bar{\d} X \; \bar{\d} Y \geq  \frac{1}{2} 
\Big| \big\langle [{\rm Re} X, {\rm Re} Y] \big\rangle \Big|,
\ee
where here
\be \label{ddd-def}
\d X := \sqrt{\langle \D X^\dag \D X \rangle}, \quad
\d' X := \sqrt{\langle \D X \D X^\dag \rangle},\quad \mbox{and}\quad
\bar{\d} X := \frac{1}{2}(\d X + \d' X). 
\ee
The proofs of these are simple. In fact, \eq{ur-g1} is a direct
application of the Schwarz inequality $| \langle \a | \b \rangle |^2 
\leq | \langle \a | \a \rangle | | \langle \b | \b \rangle |$, for
arbitrary states $\ket{\a}, \ket{\b}$. Similar inequalities can be
obtained for $ \big| \big\langle [ X^\dag, Y] \big\rangle \big|$,
 $ \big| \big\langle [ X, Y^\dag] \big\rangle \big|$,
 $ \big| \big\langle [ X^\dag, Y^\dag] \big\rangle \big|$, and the 
inequality \eq{ur-g2} is obtained by adding them together. 

For our case, the Schwarz inequality is still valid since associativity
of operator product was not needed in the proof of  it. 
Therefore we can apply, for example the uncertainty relation
\eq{ur-g2} to the commutation relation \eq{rr} and obtain the
``volume'' uncertainty relation 
\be
\bar{\d} \rh_{\bq_1}\bar{\d} \rh_{\bq_2}\bar{\d} \rh_{\bq_3} \geq
\sqrt{\frac{|\la\th_{\bq_1 \bq_2}\ra \la\th_{\bq_1 \bq_3}\ra 
\la\th_{\bq_2 \bq_3}\ra |}{8}
}
\ee
where $\th_{\bq_1 \bq_2} = (E_{\bq_1 \bq_2} + E_{\bq_1 -\bq_2}+
E_{-\bq_1 \bq_2}+ E_{-\bq_1 -\bq_2})/4$  and $E_{\bq_1 \bq_2}$ is given by  
the RHS of \eq{rr} divided by
$i$. Note however that this is purely a consequence of the commutation
relation. In our case, we have also in presence a 
nonassociative geometry with the Jacobiator relation
\be \label{XYZ}
[X,Y,Z] = i\theta.
\ee
Our goal is to extract the associated 
uncertainty relation that is a consequence
of the presence of  non-associativity. 

Writing $[X, Y, Z] = [\D X, \D Y, \D Z]$ and denote $A_{XY} := \D X \D
Y$,  $A_{YX} := \D Y \D X$ etc, we have 
\be
|\la \th \ra | \leq
| \xpv{ [A_{XY}, \D Z]} | + | \xpv{ [A_{YX}, \D Z]} | + \mbox{($X,Y,Z$
  cyclic)}. 
\ee 
Using \eq{ur-g1}, we have
\be \label{ur-na1}
\delt  A_{XY} \; \d' Z + \delt' A_{XY} \; \d Z +  
\mbox{($X,Y,Z$  cyclic)} \geq
\frac{ |\la \th \ra |}{2} 
\ee
where $\d Z$, $\d' Z$ are given by \eq{ddd-def} and 
\be 
\delt A_{XY} := \frac{1}{2} (\d A_{XY} + \d A_{YX} ), \quad
\delt' A_{XY} := \frac{1}{2} (\d' A_{XY} + \d' A_{YX} )
\ee
etc.. The relation \eq{ur-na1} gives a lower bound constraint on the
product of the coordinate uncertainties: $\d X, \d' X, \cdots $ and
of the ``area'' uncertainties: 
$ \delt A_{XY}$, $\delt' A_{XY}$, 
$\cdots$. 
Since \eq{XYZ} is invariant under the rotation group acting on $X,
Y, Z$,  it is desirable to have a form of the uncertainty relation
that is manifestly expressed in terms of $SO(3)$ invariant quantities.  
To do this, let us utilize the
inequality $ \sum x_i y_i \leq \sqrt{\sum x_i^2} \sqrt{\sum y_i^2} $
for real $x_i, y_i$ and obtain from \eq{ur-na1}
\be \label{ur-na2}
\hat{\d} R \; \hat{\d} A \geq
\frac{|\la \th \ra |}{4},  
\ee
where
\be
(\hat{\d} R)^2 := (\d_r X)^2 + (\d_r Y)^2
+ (\d_r Z)^2
\quad \mbox{with}\quad
(\d_r O)^2 := \frac{ (\d O)^2+ (\d'  O)^2}{2} 
\ee
and
\be
(\hat{\d} A)^2) := 
\frac{ (\delt A_{XY})^2 + (\delt' A_{XY})^2}{2} +
\frac{ (\delt A_{YZ})^2 + (\delt' A_{ZY})^2}{2} 
+\frac{ (\delt A_{ZX})^2 + (\delt' A_{XZ})^2}{2}. 
\ee
In the special case where $X, Y, Z$ are Hermitian, 
$\d_r O = \d O$, $\delt A_{XY} = \delt' A_{XY} $
\be
(\hat{\d} R)^2  =  (\d X)^2 + (\d Y)^2
+ (\d Z)^2
\ee
and
\be
(\hat{\d} A)^2 = (\delt A_{XY})^2 + (\delt A_{YZ})^2
+ (\delt A_{ZX})^2.
\ee
Due to their origin, we 
will call \eq{ur-na1}, \eq{ur-na2}  {\it nonassociative uncertainty
relations}. 

In our case, taking $(X,Y,Z)= (\rho_{\bq_1}, 
\rho_{\bq_2}, \rho_{\bq_3})$, let us evaluate the lower bound for 
the constraint \eq{ur-na2}. Assuming that we have monopoles of charge
$Q_\rn$ at the position $\bk_\rn = \bz_\rn$ in the BZ. It is $\sum_\rn
Q_\rn =0$
according to the  Nielsen--Ninomiya theorem. Evaluating the delta
function, we obtain in the leading order of small momentum, the
following expression for the lower bound for the
non-associative  uncertainty relations \eq{ur-na1}, \eq{ur-na2}:
\be \label{lb}
|\la \th \ra| 
= |(\bq_1 \times \bq_2) \cdot \bq_3| 
\Big|\sum_\rn Q_\rn \la N^\a_\rn\ra\Big|,
\ee
where $N^\a_\rn := N^\a_{\bz_\rn}$ and $N^\a_{\bk} = \g_{\bk}^\a{}^\dag
  \g_{\bk}^\a$ is the number operator for the orbital creation
  operator. As the expression \eq{lb}  
  depends only on simple  properties of the Weyl node, we expect to extract
  interesting information from  these uncertainty relations. This
  will be the subject for further work. 

In this paper, we have pointed out the presence of nonassociativity in the 
algebra of density operators in Weyl semimetal. This breakdown of
associativity can be thought of as some kind of ``anomaly'', with the
fundamental identity playing the role of the consistency condition. 
The  nonassociativity is supported  
by the monopoles situated at the Weyl nodes, which are also exactly
where the Fermi Arcs end. Since Fermi Arcs played a very important
role in the spectral flow between the Weyl points, it will be 
interesting to relate the algebraic
structure of the projected density operators with the properties of
Fermi Arcs and related spectral flow. It will also be interesting to 
apply the algebraic structure to  derive the  spectral sum rules for 
the density correlation functions for Weyl semimetal \cite{caza}.

\vskip7mm
\section*{Acknowledgements}
I would like to thank Daw-Wei Wang for discussions and particularly
Miguel Cazalilla for enlightening 
discussions and useful comments on the manuscript. 
This work is
supported in part by  the National Center of Theoretical Science
(NCTS) and the 
grants  101-2112-M-007-021-MY3 and 
104-2112-M-007 -001 -MY3 of the Ministry of Science and
Technology of Taiwan.


\end{document}